# A Multilayered Approach to Estimate Business Performance

Abhinov Balagoni [1], Vinay Kumar Chavala[2]

*Department of Electronics and Communication Engineering, National Institute of Technology Warangal*

[1]abhinov@ieee.org, [2] vkchavala@gmail.com

*Abstract*-The performance of a business is chiefly determined by the business model adopted and its environment of operation. This paper emphasizes on the integration of Web 2.0 services like social networking to your existing business, which helps in the progressive sustainability of the business organization. The usage of Web 2.0 tools, such as Wikipedia, blogs and social networking services such as Facebook, LinkedIn, Orkut by individuals in all societies, has been pervasive and very successful in proliferating their use at the professional or business levels. This paper puts forth the discussion which helps in better understanding of what social networking encompasses for present day business. It also aims to educate business decision makers about the benefits and risks associated in incorporating business model with social networking. Thus, the paper focuses on the implementation of social networking platforms in business operations and discusses the different attributes of business performance and gives an overview in choosing a social networking platform for its business purposes.

*Index Terms*-Business Performance, Social Networking, Adaptability

## I. INTRODUCTION

The performance of a business can be determined chiefly by its business model and the environment in which it operates [1].

The term business model has a broad scope of definitions, for example, scope of definitions for a business model may include the basic definition by [2] as the method of doing business by which a company can sustain itself and do its business operations [3] - that is, generate revenue.

The environment, in which the business operates, includes factors beyond the control of the business, affecting its functioning. These include social, political and technological factors etc. While some of these factors may have direct influence on the business performance, others may influence indirectly. Thus, business environment may be defined as the set of external factors, which have a direct or indirect bearing on the functioning of business. In last few decades, social networking platforms have evolved rapidly in terms of enabling knowledge sharing, collaborative efforts among knowledge sharing, collaborative efforts among individuals and organizations. In the organizational context, it can help us in hiring the right people and marketing the product. Terms such as business networks [4], supply chains [5] etc., clearly imply that organizations will need to function cooperatively to optimize their performance. Thus, social networking is an indispensable tool for a business nowadays but the difficulty lies in applying it for optimizing business performance. The role that social networking can play to foster such cooperative organizational environments is an important and exciting area of study.

## II. MULTILAYERED APPROACH

The pyramid, in the figure 1, shows an overview of

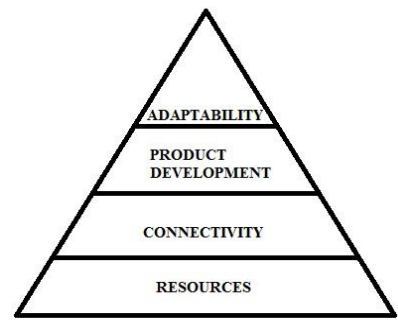

Figure1: Pyramid Layered Approach

generic attributes of a business performance. It has been synthesized based on a thorough analysis of the body of literature about business models [6] [7]. In order to provide a framework for a systematic analysis of the performance of a business, the analysis of pyramid layered approach is developed. The starting point for the development of the framework was the most cited specific definition proposed by Timmers. According to Timmers, a business model is an architecture for the product, service and information flows, including a description of the various business factors and their roles. A description of the potential benefits for the various business factors. A description of the sources of revenues [8]. The components denoted by Timmers definition were extracted and enhanced with further aspects affecting business and is consolidated to form the pyramid layered structure as shown in figure 1. The pyramid layered approach discussed here segregates an operational business process into four levels viz. resources, connectivity, product development and adaptability. The different layers of the pyramid layered structure are defined below.

The resources layer reflects the inputs required for the operation of a business, in other words it can be called a foundation stone for a business process. The connectivity layer features the interaction of resources among themselves and estimates their outcome. The product development layer covers the aspects of business development like collaboration and sustainability of the business.

The adaptability layer identifies the new trends in the business and imbibes it to update and make the business operation more efficient.

Social networking, an aspect of the internet technologies, has undergone numerous changes and altered into an indispensable tool for business these days. Social networking is fundamentally shifting the way we communicate, it is modifying conventional boundaries, increasing transparency in everything we do. Thus, social networking platforms help in connecting a major section of society and hence companies, large and small, can no longer ignore or try to block social networking platforms in their environment. The basic motto is to reach target audiences and people are more likely to participate in a social media forum than any other venue in coming future. Customers, partners, and employees, alike expect to engage with us via social media. It is a means to stay connected, collaborate, recruit and gather feedback.

As a result, we need to support social media in our business environment to enhance the innovation, productivity, and accelerate growth which essentially drive business. As a means of winning new opportunities, the usage of social networking platforms by business continues to expand. According to a global survey of business use of social networks by regus,7 percent more companies have discovered new business clients through social networks compared to Q2 2010(up from 40 percent in July 2010) [9] and also reported commercially successful. A major proportion of these companies have reported higher income and profits when compared to those that do not use social networks to acquire new business. The section III discusses in detail the various requisites of business and the resources available on different social networking platforms.

Integrating the social networking aspect to your business helps to connect a huge part of the customer base. Social networking gives your business a way to appeal to customers and draw them in. Nowadays, many people find business information on the social networking sites than they do in the phone book. In this technological age, there is no reason not to use social networking to help your business.

III. DETAILED DESCRIPTION OF COMPONENTS

*A. Resources*

The pyramid layered approach defines resources as the tools for business and professionals of similar areas of interest and background by bringing them to a common platform to share and contribute vital professional information for career advancement of a professional and human resource management of a business organization. In short, resources layer in the pyramid layered approach concentrates services for personal professional development at the human resources level. It is a pool for the management and human resource professionals to develop their skill and search for skilled professionals to accomplish a job.

Table I: (A) Comparison for the attributes of resources. NA - Not Available, A - Available

| Parameter | Resume Uploading | Networking Groups | Web Search ability | Poll |
|---|---|---|---|---|
| Facebook | NA | A | A | A |
| LinkedIn | A | A | NA | A |
| Google plus | NA | A | NA | A |
| My space | NA | A | NA | A |
| Orkut | NA | A | NA | A |
| Twitter | NA | A | NA | A |
| Viadeo | A | A | NA | A |

Table I: (B) Ranking of social networking services. NA - Not Available, A - Available

| Parameter | Alexa Global Rank | Compete US Rank | Quantcast US Rank |
|---|---|---|---|
| Facebook | 2 | 2 | 2 |
| LinkedIn | 12 | 44 | 20 |
| Google plus | NA | NA | NA |
| My space | 155 | 49 | 79 |
| Orkut | 174 | 5449 | 14,198 |
| Twitter | 9 | 23 | 6 |
| Viadeo | 441 | 21367 | 5017 |

The table I(A) shows an overview of the different attributes of resources like a) resume uploading b) networking groups c) web search d) polls in various social networking sites and table I(B) shows the global page ranking given by Alexa [10], and United States page rank by Compete [11] and Quantcast [12].The social networking sites, LinkedIn and Viadeo, cater the professional needs to a more extent when compared to other platforms.

*B. Connectivity*

The next layer of pyramid layered approach focuses in the connectivity among resources. The role of connectivity is to exchange information between businesses or professionals within a business, in the proper format and appropriate timeframe. Synchronous and asynchronous are the two major modes of such communication; synchronous communication signifies the communication carried out between two connected entities at the same instant and asynchronous communication signifies the communication, where the sender and transmitter need not be on at the same instant of time. A sender might send information and the system will store it for the receiver to pick up at a later time. Thus, we may also categorize connectivity as unicast communication, where there is a single sender and a single receiver, versus a broadcast communication, where a single sender may send to one or many receivers.

Table II: Comparison of the attributes of connectivity.

| Parameter | Instant Messaging | Email | File Sharing | Voice Calling | Video Calling | Threaded Discussion Forum |
|---|---|---|---|---|---|---|
| Facebook | A | A | A | A | A | NA |
| LinkedIn | NA | NA | A | NA | NA | A |
| Google plus | A | NA | NA | A | A | NA |
| Myspace | A | NA | A | NA | NA | NA |
| Orkut | A | NA | NA | NA | NA | NA |
| Twitter | NA | NA | NA | NA | A | A |
| Viadeo | NA | NA | NA | NA | NA | A |

Lastly, we can categorize connectivity further as to whether the conversation is to be recorded or not or whether the conversation are intended to be retained for a period of time or made permanent. The table II gives the analysis of the attributes of connectivity like a) instant messaging b) email c) file sharing d) voice calling e) video calling and f) threaded discussion forum of different social networking platforms associated with their global page ranking given by Alexa, and United States of America page rank given by Compete and Quantcast. Thus, we can compare the social networking services that suit the requirements of business organization. Among the social networking sites, Facebook provide a better means to connect when compared to other platforms as can be seen in the table.

*C. Product Development*

The product development layer comprises of integration of processes in a business organization and is concerned with the privacy issues of the professionals and also with the security levels of the information shared.

*1) Collaboration*

The production of a business plays a vital role in its success and its efficiency can be increased by business process integration. Integration is the process of merging different technologies to create a single technology that can either replace or updates the original technologies. Business process integration [18] is a business strategy [15] to link individual technology. It enhances the efficiency, but the major problem with the BPI is how to integrate both the technologies. For example, Twitter integrated with yahoo [17], Google, Microsoft in 2010, this integration will include a Twitter feed in several Yahoo services and the ability to update your Twitter status from within Yahoo. Similarly Facebook's Frictionless sharing [19] integration with Yahoo! News US resulted in further integration of with 26 new sites like Yahoo! News Philippines and France and about 1 million articles are being shared on Facebook [16]. Facebook integrates with Skype [19] to introduce video calling that lead to an increase in its market value.

Facebook collaborated with Bharti Airtel which enabled about 130 million plus customers across India to access Facebook's full mobile site at no data cost in 2010. It was a great move by Facebook against its opponent Orkut in India as the search trend for Facebook is steady and stronger during 2010. Figure 3 and figure 4 shows the comparison of traffic of Facebook and Orkut in 2010 during which Facebook is dominating over Orkut in India. Bharti Airtel was successful in inculcating the habit of data consumption to its customers and once people are hooked on to the habit of data consumption, they will automatically subscribe the GPRS pack and hopefully aid in increasing the ARPU [14] of the Bharti Airtel. Thus it clearly tells us that collaboration results in mutual benefit of the organizations either in gaining market share, reducing cost and improving quality [13].

*2) Privacy and Security Issues*

Spread of information is quicker through social networking platforms when compared to conventional network but the widespread of information on social networking sites can be supported only when they are secured and can be limited from others. Most of the social networking sites allow the third party developers to develop applications which require the profile information of the users causing a privacy risk for users.

Facebook and Google platforms allow the third party developers to launch their applications on their respective social networking sites [24]. Facebook third party application developer policies [20] clearly states that applications used by their users have access to their profile information to operate properly. These applications have access to the public information of the users available in their profile and also to the information that users have made visible" for all" through privacy preferences of their profile. Thus users of social networking sites are required to give full read permissions to all profile information to operate the applications [21].

The Figure 2 shows that application can be added by user if and only if he/she grants full access to his profile, as well as to the profile data of his friends. If user doesn't agree to share his profile information, there is no grant of full access to the application. As a result third party developers have the profile information of the users without the application need [21] [22] [23].

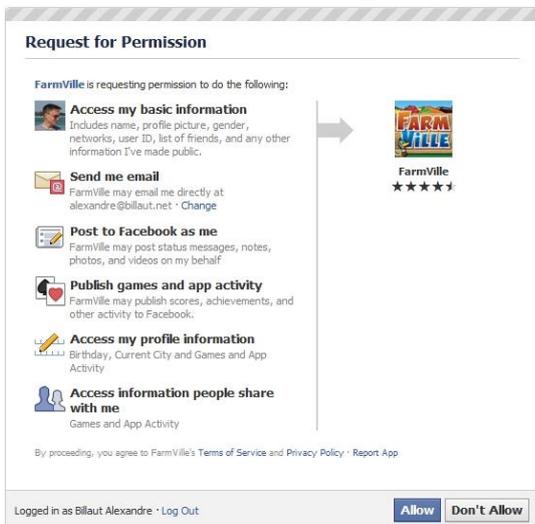

Figure 2: Shows how an application will access profile information

### D. Adaptability

Every business within a certain industry and market, has certain threats like competitors. To stay healthy and sustain over time, the business must be adapting itself to the ever changing demands of its ecology. As the environment changes, so must the business incorporate changes to meet the new demands.

Social networks often act as a relationship management tool for companies selling products and services. Companies can also use social networks for advertising in the form of ads, makes easy to keep in touch with the contacts of the business organization around the world. For the organizations to extract maximum benefits, the social networking platform opted by them must be in line with the present trends.

Checkpoints of adaptability are Search traffic and Number of registered users.

Search traffic, number of registered users depicts the present trend of the people and higher these parameters value, more suitable it is for an organization to work in that particular social networking. For example if you consider figure 3 and figure 4 shown below,

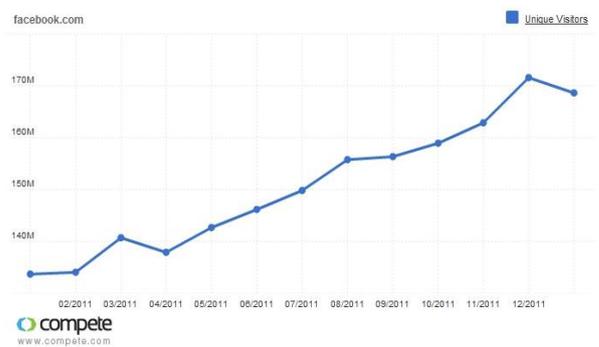

Figure 3: Search traffic of Facebook

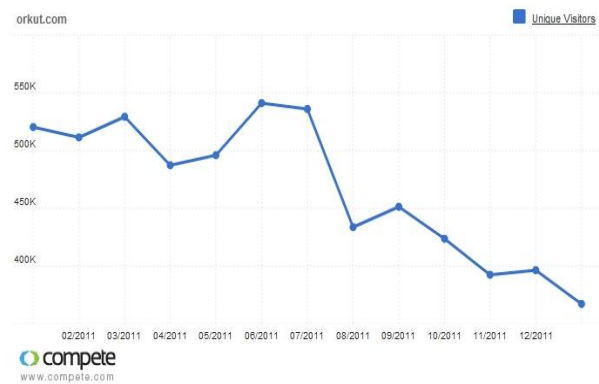

Figure 4: Search traffic of Orkut

We can see that unlike in Facebook, the search traffic is decreasing in the Orkut. Hence for a business organization it is not advisable to invest in Orkut (decreasing graph) given the present day scenario.

An example to state the importance of adaptability in a business, In India, Facebook had approximately 0.7 million users and whereas Orkut had 3 million active users during 2008 but now Facebook has around 43.4million users in India. This clearly states that business must adapt according to the present day conditions so as to develop progressively and sustain the competition. The Table III shows a comparison of these parameters to the various social networking sites.

Table III: Comparisons of various parameters

| Parameters | Search Traffic | Number of Registered Users |
|---|---|---|
| Facebook | 171.58M | 800M |
| LinkedIn | 24.44M | 135M |
| Orkut | 396.37K | 66M |
| Twitter | 40.41M | 300M |
| Xing | 80K | 3.88M |
| Ryze | 4.92K | 5 Lakhs |
| My space | 21.06M | 30M |
| Google plus | 20.03Lakhs | 90M |

M-Million, K-Thousand

IV. CONCLUSION AND FUTURE SCOPE

This paper increases the understanding of what social networking encompasses and how it can be utilized for business purposes. The ideas and discussion put forth may give new insights into the social networking applications as part of the services provided by businesses. We focussed on how we can effectively use the power of social networking technologies to enhance a business performance through a multi-layered approach, and in that process suggest the social networking platforms to optimize the business operations.

The applicability of social networking platforms for business performance enhancement is still in the budding stage. Integrating social networking technologies with your existing business helps in leveraging the business operations and enhances its performance with limited resources. However, the concerns regarding the privacy of the conversations of the professionals and the security of the information shared on social networking platforms is not up to expected standard.

The future scope of study of this work may include the development of a mathematical model, where in the attributes of different layers of pyramid model may be given appropriate weights, which may estimate the performance of a business accurately.